\documentclass[pra,twocolumn,showpacs,aps,floatfix,amsmath,amssymb,amsfonts]{revtex4-1}

\usepackage{amsmath}
\usepackage{graphicx}
\usepackage{color}

\graphicspath{{figures/}}

\newcommand{\beq}{\begin{equation}}
\newcommand{\eeq}{\end{equation}}

\begin{document}

\title{Impact of overlapping resonances on magnetoassociation of cold molecules in tight traps}
\author{Krzysztof Jachymski$^{1,2}$}
\affiliation{
$^1$ Faculty of Physics, University of Warsaw, Pasteura 5, 02-093 Warsaw, Poland\\
$^2$ Institute for Theoretical Physics III, University of Stuttgart, Pfaffenwaldring 57, 70550 Stuttgart, Germany}
\date{\today}

\begin{abstract}
Overlapping Feshbach resonances are commonly observed in experiments with ultracold atoms and can influence the molecule production process.
We derive an effective approach to describe magnetoassociation in an external trap in the presence of multiple overlapping resonances. 
We study how the strength and shape of the trap affects the energy level structure and demonstrate the existence of a regime in which the conventional two-channel Landau-Zener description of the molecule production process breaks down. 
\end{abstract}

\maketitle
\section{Introduction}
Ultracold molecules are a promising platform for engineering exotic many-body hamiltonians, investigating the physics of strongly dipolar bosons and fermions and performing quantum computations~\cite{Carr2009}. However, production of a large sample of ultracold molecules with low entropy is an extremely challenging experimental task~\cite{Ni2008,Molony2014,Takekoshi2014,Deiss2014,Park2015}. The most successful schemes to date rely on association of molecules from ultracold atoms rather than direct cooling~\cite{PSJ2012}, although notable progress has recently been made with direct methods as well~\cite{Prehn2016,Norrgard2016}. The association process starts with two overlapping atomic clouds, from which the molecules can be produced by magneto- or photoassociation. A resonance between the atomic pair and a molecular state is needed for efficient conversion. Magnetic Feshbach resonances are commonly used at this step~\cite{JulienneRMP,Mies2000,Julienne2004}. By slowly changing the magnetic field, pairs of atoms are adiabatically transferred to the most weakly bound state. These weakly bound molecules can then be converted to deeply bound states using Stimulated Raman Adiabatic Passage (STIRAP) technique.

Feshbach resonances allow not only for production of cold weakly bound molecules, but are one of the crucial tools used for controlling the interactions in an ultracold gas~\cite{JulienneRMP}. Identification of Feshbach resonances becomes then an essential step in experiments with new species~\cite{Julienne2009,Takekoshi2012,Berninger2013}. It is important to find resonances with magnetic field width big enough for precise experimental control. Methods for engineering the resonance width with external electromagnetic fields can thus be useful, especially for closed-shell atoms which generally do not exhibit wide resonances~\cite{Tomza2014}.

A great improvement of the conversion efficiency and control can be gained by using an optical lattice to confine the atoms before the Feshbach ramp~\cite{Rom2004,Volz2006,Lang2008}. Ideally, one wants to prepare a system in which each lattice site is occupied by exactly two atoms. High densities and low entropy can be obtained by starting from the Mott insulator state\cite{Moses2015,Covey2015}. This requires careful planning, taking into account both inter- and intraspecies interactions~\cite{Naini2015}. For deep lattices, a single site can be well approximated by harmonic oscillator potential. Resonances in harmonic traps are theoretically well understood~\cite{Bolda2002,Blume2002,Idziaszek2005}. However, for association of heteronuclear molecules the trapping potential is no longer purely harmonic, but contains a term that couples the center of mass and relative motion of the atomic pair~\cite{Sengstock,Bertelsen2007,Jachymski2013a}. Other anharmonic corrections may arise when the lattice potential is not very deep.  

In many currently studied cases, from cesium dimers~\cite{Berninger2013} to systems involving lanthanide atoms~\cite{Petrov2012,Martinez2015} one deals with a multitude of overlapping resonances which may be hard to separate~\cite{Jachymski2013b}. While Feshbach association in these systems is still possible~\cite{Frisch2015}, it requires a lot of care as one may associate many different states during the process~\cite{Covey2015}. Optimal control techniques~\cite{Glaser2015} might be used here to avoid unnecessary products. However, to design a well-shaped control pulse, it is necessary to know the energy level structure quite accurately. Densely overlapping resonances are also expected to be ubiquitous in molecular systems~\cite{Mayle2013}, which can have consequences for their many-body dynamics~\cite{Doccaj2015}. 

A convenient method of describing a single resonance in an isotropic trap, based on a two-channel configuration interaction model, has been developed in~\cite{Jachymski2013a}. By expanding the anharmonic terms in the harmonic oscillator basis, it has been possible to make use of its analytic properties to efficiently renormalize the resonance shifts and obtain the system eigenstates. In this work, we generalize these results to the case of multiple overlapping resonances in anisotropic potentials, allowing to describe most current experiments studying cold molecule association. 

The paper is structured as follows. In Section II, we derive the multichannel model describing in principle arbitrary trapping potential. In Section III we apply the formalism to the case of an anisotropic harmonic trap and discuss the renormalization method for this specific case and the connection of the model parameters to physical quantities. In Section IV we study several exemplary cases, showing under what conditions the overlapping resonances can be separated. In Section V we discuss the limitations and possible extensions of the model.

\section{Feshbach resonances in a trap}
To describe Feshbach resonances (FR) in an external trap, we will use a simple model with the couplings between open and closed channels replaced by Dirac delta terms with the parameters chosen properly to reproduce the true resonance parameters (we focus on $s$-wave interactions between identical bosons or distinguishable particles). This pseudopotential approximation can be justified as long as the characteristic length of the true interaction potential is much smaller than the trap length scales and allows for simple description of resonances regardless of their exact nature such as the bound state wave function, which can in principle be a higher partial wave bound state, and structure of short-range couplings. Two-channel contact interaction models have been very successful in describing the properties of ultracold bosons and fermions~\cite{Gurarie2007,Buchler,BlochRMP}. However, they require renormalization of the resulting ultraviolet divergences. For a single channel, short-range interaction can be conveniently described by regularized Dirac delta $V(r)=\frac{4\pi \hbar^2 a}{m}\delta(\mathbf{r})\frac{\partial}{\partial r}\left(r\cdot\right)$~\cite{Wodkiewicz1991} with $a$ being the scattering length. For two particles in a trap, a self-consistent formula for the energy levels can be found using such interaction potential~\cite{Bush,Idziaszek2005}. Tight traps require using energy-dependent scattering length to obtain reliable results~\cite{Bolda2002}. Applying a two-channel model, which is necessary for describing the association process in the case of a narrow resonance, results in formulas with similar structure as the single-channel case. This provides a very convenient renormalization scheme~\cite{DienerHo2006,Jachymski2013a} and allows for matching the coupling strength between the channels with Feshbach resonance parameters. Generalization of this method to multiple closed channels is the scope of the present section.

Let us assume that different closed channels are not coupled to each other (they have been pre-diagonalized). For simplicity, we will also neglect the structure of the molecular states and background interactions in the open channel. These assumptions allow to reproduce correctly the properties of the most weakly bound states, which are precisely the states we are interested in. Consequently, the hamiltonian can be written as
\beq
\begin{split}
H=\left|o\right>\left<o\right|\left(\frac{p^2}{2\mu}+\frac{P^2}{2M}+V(\mathbf{r},\mathbf{R})\right)+\\
+\sum_{i}{\left|\chi_i\right>\left<\chi_i\right|\left(\frac{P^2}{2M}+\tilde{V}(\mathbf{R})\right)}+\\
+\sum_{i}{\left(\left|\chi_i\right>\left<o\right|W_i+\mathrm{h.c.}\right)},
\end{split}
\label{eq:ham}
\eeq
where $\left|o\right>$ labels the open channel, $\left|\chi_i\right>$ denotes the closed molecular channels, $\mu$ is the reduced mass of the pair, $M$ is the total mass, $p$ describes the relative momentum and $P$ the center of mass momentum. Furthermore, $W_i(\mathbf{r})=g_i\delta(\mathbf{r})$ describes the couplings and $V$ is the trapping potential. The $\tilde{V}(\mathbf{R})$ potential is obtained as $V(0,\mathbf{R})$. The trapping potential can in principle be different for different molecular channels but we neglect this here, as the weakly bound Feshbach molecules have generally very similar polarizabilities and thus see the same trapping field.

The wave function describing the eigenstate of this hamiltonian can be decomposed into single channel eigenstates
\beq
\label{eq:wavef}
\left|\Psi\right>=\left|o\right>\sum_{n}{c_{n}\psi_n(\mathbf{r},\mathbf{R})}+\sum_i{\left|\chi_i\right>\sum_{p}{a^i_p \Phi_p(\mathbf{R})}},
\eeq
where the $\psi_n$ and $\Phi_N$ fulfill
\begin{equation}
\left(\frac{p^2}{2\mu}+\frac{P^2}{2M}+V(\mathbf{r},\mathbf{R})\right)\psi_n (\mathbf{r},\mathbf{R})=\epsilon_n \psi_n (\mathbf{r},\mathbf{R})
\end{equation}
\begin{equation}
\left(\frac{P^2}{2M}+\tilde{V}(\mathbf{R})\right)\Phi_{p} (\mathbf{R})=(\nu_i(B)+\varepsilon_p) \Phi_{p} (\mathbf{R}).
\end{equation}
We note that here $\epsilon$ denotes the eigenenergies of the open channel, while $\varepsilon$ is used for closed channels. We use the same wavefunctions $\Phi$ for each closed channel as the trapping potential felt by the molecules is the same. The only difference between the closed channels are different resonance shifts $\nu_i(B)$.

Inserting ansatz~\eqref{eq:wavef} into the Schr\"{o}dinger equation leads to a set of equations for the expansion coefficients which can be projected onto single basis states thanks to the orthonormality of the basis set. This gives
\begin{eqnarray}
(E-\epsilon_n)c_n&=\sum_{i}{\sum_k{a_k^i V_{kn}^i}}
\label{eqc}
\\
\label{eqa}
(E-\varepsilon_p-\nu_i(B))a_p^i&=\sum_s{c_s \left(V_{ps}^i\right)^\star},
\end{eqnarray}
where $V_{kn}^i=g_i\left<\psi_n(\mathbf{r,R})|\delta(\mathbf{r})\Phi_k(\mathbf{R})\right>$. We can now insert the $a$ coefficients obtained from~\eqref{eqa} into~\eqref{eqc}, arriving at
\beq
(E-\epsilon_n)c_n = \sum_i{\sum_{kj}{\frac{\left(V_{kj}^i\right)^\star V_{kn}^i}{E-\nu_i-\varepsilon_k}c_j}}.
\label{eq:selfcons}
\eeq
This equation in general requires renormalization. For a two-channel contact interaction model it is well-known how to treat the ultraviolet divergence~\cite{Gurarie2007,Kestner2007}. Due to our assumption of uncoupled molecular states, each closed channel can be renormalized separately, reducing the problem to the two-channel case. It is also possible to derive a general renormalization scheme for the case of coupled closed channels~\cite{Doccaj2015}. In the next section we will apply a slightly different scheme which makes use of the analytical properties of the harmonic trap eigenstates~\cite{Idziaszek2005,Jachymski2013a,DienerHo2006}.

\section{Anisotropic harmonic traps}
One of the simplest and widely applicable examples of the trapping potential is the harmonic one. In this section we will show how the formalism works for harmonic confinement alone, but adding anharmonic terms which couple center of mass and relative motion is straightforward in the present model, as one can diagonalize the anharmonic terms in the harmonic oscillator basis. A common example of a simple anharmonic term occurs for association of heteronuclear molecules. In this case, the atoms forming a pair feel different trapping frequencies and a coupling term of the form $C \mathbf{R\cdot r}$ appears in the open channel. 

For harmonic confinement the open channel wavefunctions can be written as $\psi_n(\mathbf{r,R})=\phi_{s(n)}(\mathbf{r})\Phi_{t(n)}(\mathbf{R})$ (here the index $n$ has been split into two parts describing the relative and center of mass degrees of freedom) and the coupling $V_{kn}^i$ reduces to $g_i \phi_s(0)\delta_{kt}$. In a harmonic trap we can then restrict the problem to a single bound state in each channel as the center of mass is not affected. For practical reasons, the traps used in experiments are often anisotropic, so we assume that the trapping potential has the form
\beq
V(\mathbf{r})=\frac{1}{2}\mu\omega^2(z^2+\eta^2 \rho^2)
\eeq
with $\eta>0$ being the trap anisotropy. For simplicity of the notation we will only consider a single closed channel in this section, as the generalization to many channels has been provided above. Furthermore, we will measure the energy in units of $\hbar\omega$ and the lengths in units of the harmonic oscillator length $a_{ho}=\sqrt{\hbar/\mu\omega}$. The wave function can be expressed in the harmonic oscillator basis as
\beq
\label{eq_wavefharm}
\begin{split}
\left|\Psi\right>=\left|o\right>\sum_{k,K,n,N}{c_{nNkK}\phi_n(\rho)\Phi_N(\varrho)\upsilon_k(z)\Upsilon_K(Z)}+\\+\left|\chi\right>\sum_{ij}{a_{ij}\Upsilon_i(Z)\Phi_j(\varrho)},
\end{split}
\eeq
where indices $n,k$ mark the wavefunctions describing the relative motion, and $N,K$ the center of mass motion and $\phi$, $\Phi$, $\upsilon$ and $\Upsilon$ are the eigenfunctions of the 2D and 1D harmonic oscillator, respectively. Inserting this into the Schr\"{o}dinger equation, we get
\beq
\label{eq_an}
(\varepsilon_{pq}-\nu-E)a_{pq}+\sum_{ijkKnN}{\frac{V_{kKnN}^{pq}V_{kKnN}^{ij}}{\epsilon_{nk}+\varepsilon_{NK}-E}a_{ij}}.
\eeq
Here, as previously, $\varepsilon$ denotes the energy of the center of mass motion eigenstate, while $\epsilon$ is used for relative motion. The coupling matrix elements are given by
\beq
V_{kKnN}^{pq}=g\delta_{pK}\delta_{qN}\phi_n(0)\upsilon_k(0)=g\sqrt{\frac{\eta}{\pi}}\frac{2^k \pi^{1/4}}{\sqrt{2^k k!}\Gamma\left(\frac{1-k}{2}\right)}.
\eeq
We note that as the 2D harmonic oscillator wavefunctions have the same value at the origin, the coupling term is independent of $n$. Eq.~\eqref{eq_an} reduces then to
\beq
\label{eq_anf}
\mathcal{E}-\nu=\frac{g^2\eta}{\sqrt{\pi}}\sum_{nk}{\frac{2^k}{k! \Gamma\left(\frac{1-k}{2}\right)^2}\frac{1}{2\eta n+k-\mathcal{E}}}.
\eeq
Here $\mathcal{E}=E-\varepsilon_{pq}$. 

It possible to directly sum eq.~\eqref{eq_anf} over $k$, arriving at
\beq
\mathcal{E}-\nu=\frac{g^2 \eta}{2\pi}\sum_{n}{\frac{\Gamma\left(\eta n-\mathcal{E}/2\right)}{\Gamma\left(\eta n-\mathcal{E}/2+1/2\right)}}.
\eeq
We now notice that summation up to $n^\star$ results in a divergent term $(2\sqrt{n^\star/}+\zeta(1/2))/\sqrt{\eta}$ with $\zeta$ being the Riemann zeta function. We can extract this divergence by subtracting $1/\sqrt{\eta(n+1)}$ under the sum~\cite{Idziaszek2006,DienerHo2006}. Then we define the renormalized resonance shift $\nu^\star$ as $\nu^\star=\nu-\frac{g^2 \sqrt{\eta}}{\pi}\sqrt{n^\star}$, which removes the divergence from the calculations. 

One can also look at the problem from a formally different way, introducing integral representation of eq.~\eqref{eq_anf} via the substitution~\cite{Idziaszek2006}
\beq
\frac{1}{2\eta n+k-\mathcal{E}}=\int_{0}^{\infty}{dt\,e^{-t(2\eta n+k-\mathcal{E})}}.
\eeq
This allows to perform the summations analytically and obtain
\beq
\label{eq:selfI}
\mathcal{E}-\nu=\frac{g^2 \eta}{\sqrt{\pi}}\int_{0}^{\infty}{dt\frac{e^{t\mathcal{E}}}{(1-e^{-2\eta t})\sqrt{1-e^{-2t}}}}.
\eeq
The divergence which is present in formula~\eqref{eq:selfI} is exactly equivalent to the divergence of the two-particle Green function for a single-channel problem with delta interactions in free space. The detailed treatment of this type of divergence is given in~\cite{Idziaszek2006}. Briefly, the short-range limit $\mathbf{r}\to 0$ of this integral, where the divergence should occur, is dominated by small argument $t$. One can check that in the leading order for small $t$ the integral gives $1/2\pi r$. This term in the single channel model can be removed using regularization operator $\frac{\partial}{\partial r} (r\cdot)$. The presence of the trap is not important here, as the trapping potential can be regarded as constant in the interaction region. Formula~\eqref{eq:selfI} can thus be renormalized by subtracting $1/t^{3/2}$ term under the integral, which effectively removes the divergent part of the free space Green function and recovers the single-channel result of~\cite{Idziaszek2006}.

An important step now is to make a connection between the parameters $g$, $\nu^\star$ and the Feshbach resonance parameters~\cite{JulienneRMP} $s_{res}=a_{bg}\Delta\delta\mu$, where $a_{bg}$ is the background scattering length, $\Delta$ is the magnetic field width and $\delta\mu$ is the magnetic moment difference between the open and closed channel states. This can be done by taking $\omega\to 0$ limit at constant energy and comparing with the free space expressions for the bound state energy and scattering length~\cite{Jachymski2013a,DienerHo2006}. This yields (back in SI units)
\beq
\nu^\star=\delta\mu(B-B_0),
\eeq
where $B_0$ is the resonance position, and
\beq
g/(\hbar\omega)=\sqrt{\frac{2\pi a_{bg}\Delta\delta\mu}{a_{ho}\hbar\omega}}.
\eeq

For a set of overlapping resonances, each of the can be characterized by its own $\Delta_i$, $\delta\mu_i$ ($a_{bg}$ can usually be taken as constant) and then renormalized separately, as pointed out in the previous section. The results are valid for arbitrary $\eta$ as long as the shortest trap lengthscale is still much larger than the interaction range. One can also introduce additional terms to the trapping potential that couple center of mass and relative motion and look for the energy levels in the same way. The only difficulty lies then in expressing the open channel eigenstates in harmonic oscillator basis, the structure of eq.~\eqref{eq_an} remains similar~\cite{Jachymski2013a}.

\section{Examples}
\begin{figure*}
\includegraphics[width=0.4\textwidth]{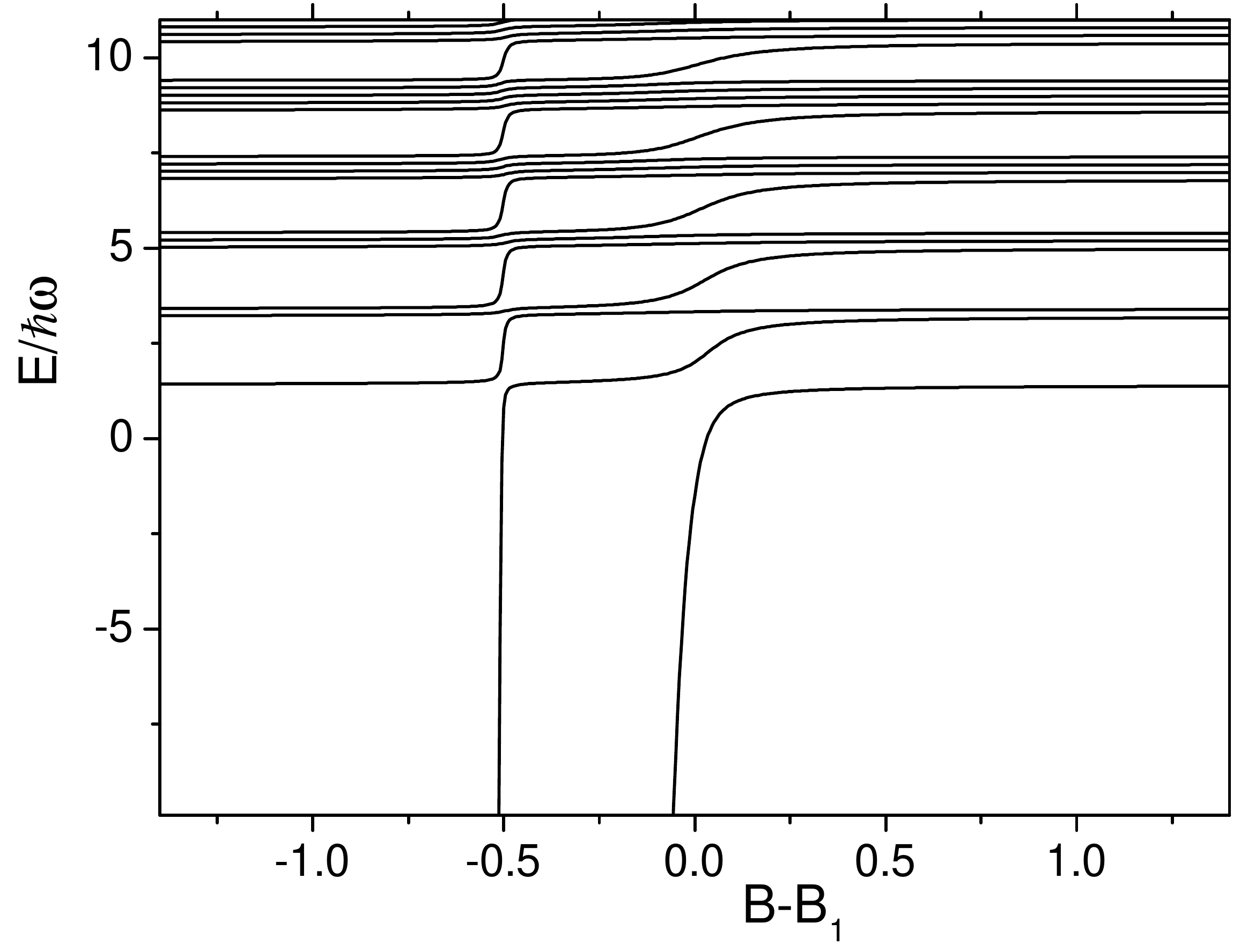}
\includegraphics[width=0.4\textwidth]{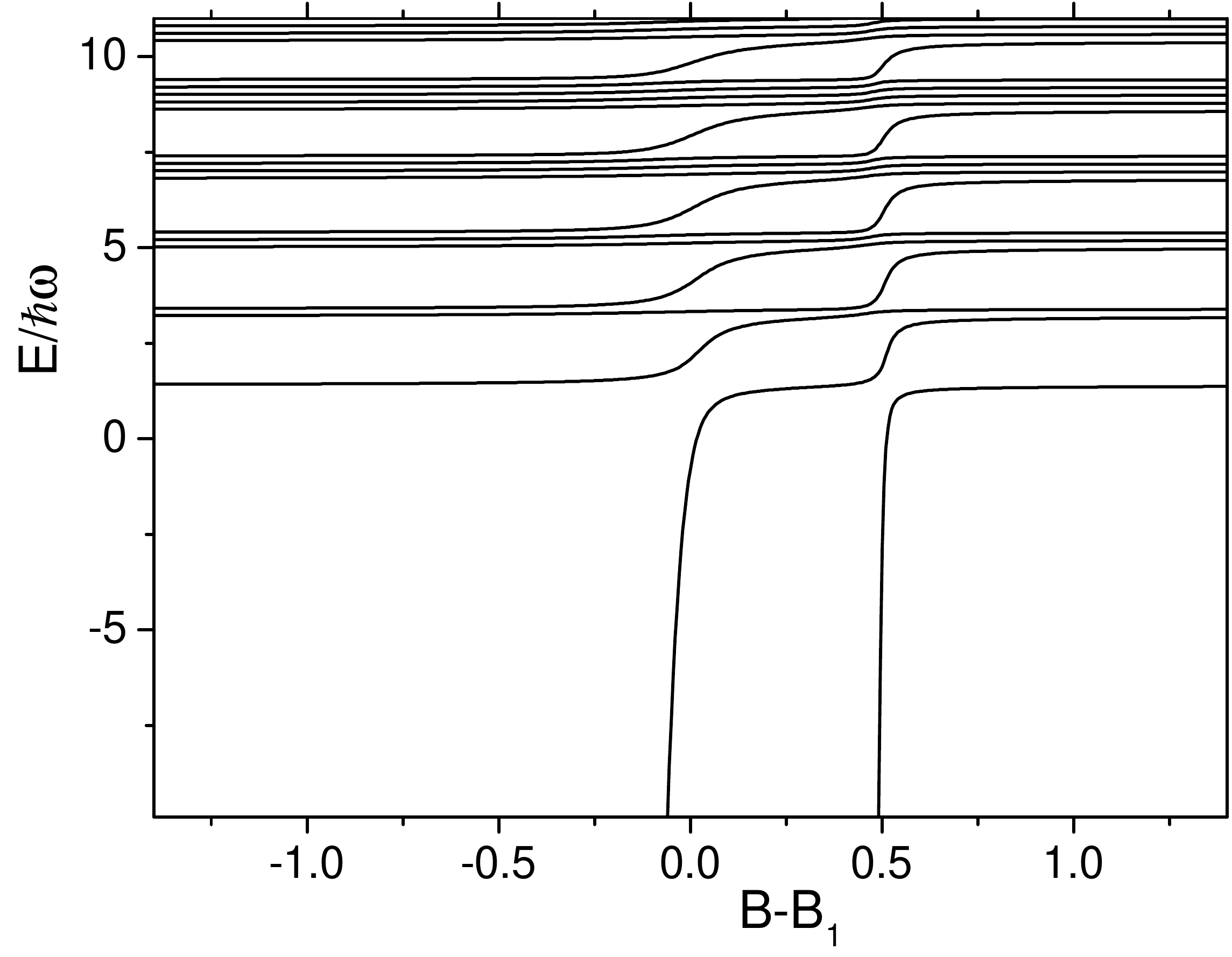}
\caption{\label{fig1} Role of the relative positions of resonances. Figure shows the energy levels for two resonances separated by 0.5 G~\cite{GaussNote} in a 1 kHz trap with slight anisotropy $\eta=0.9$. The stronger resonance has the width $\Delta=3$ G and the weak one $\Delta=0.5$ G. Left: the weak resonance is located on the left of the strong one, right: the weak resonance is located at higher magnetic field. The observed width of the weaker resonance is much larger for the latter case. Magnetic field B is given in units of gauss in all the figures throughout and $B_1$ always marks the position of the resonance occurring at smaller field.}
\end{figure*}

We will now analyze several experimentally realistic examples, showing possible consequences of the presence of overlapping resonances. Let us first focus on the case of a weak resonance located in the vicinity of a much stronger one. It is a very common case in alkali atoms that a broad $s$-wave resonance is accompanied by a narrow one, for example coming from weak coupling of the entrance channel to a $d$-wave bound state~\cite{Berninger2013,Julienne2009}. Then the broad resonance is mostly unaffected by the other one and can be thought of as setting the local background scattering length. In free space, this results in modification of the observable width of the narrow resonance depending on the relative position of the two resonances~\cite{Jachymski2013b}. This behavior is also visible in the trapped case. Figure~\ref{fig1} illustrates this for the case of two atoms with rubidium mass and $a_{bg}=100a_0$, where in one case the narrow resonance occurs at lower magnetic field (left), and in the other case it occurs at higher field (right), but with the same difference. It is evident that the narrow resonance can be strongly influenced, which manifests itself as stronger or weaker anticrossings of the energy levels.

\begin{figure*}
\includegraphics[width=0.4\textwidth]{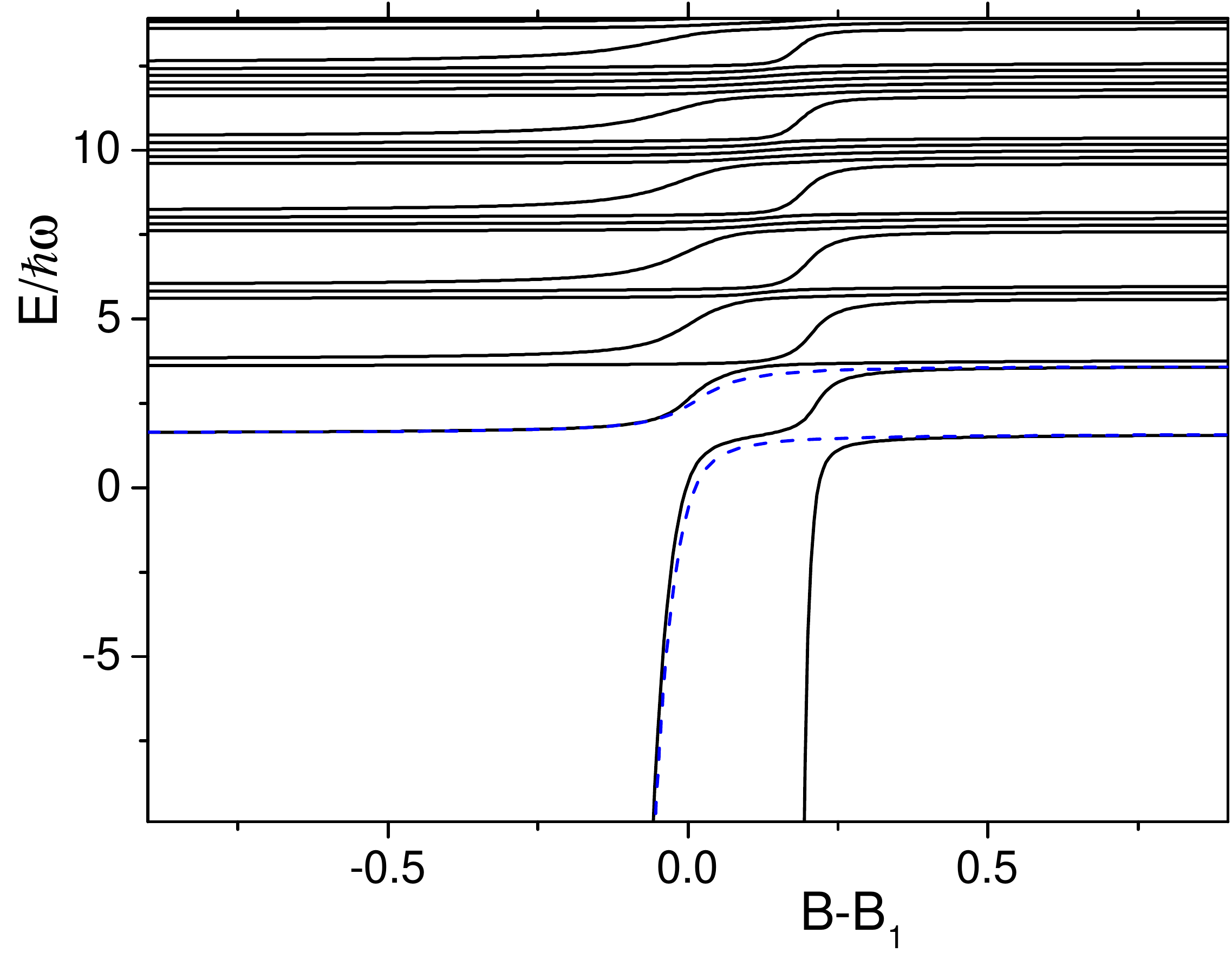}
\includegraphics[width=0.4\textwidth]{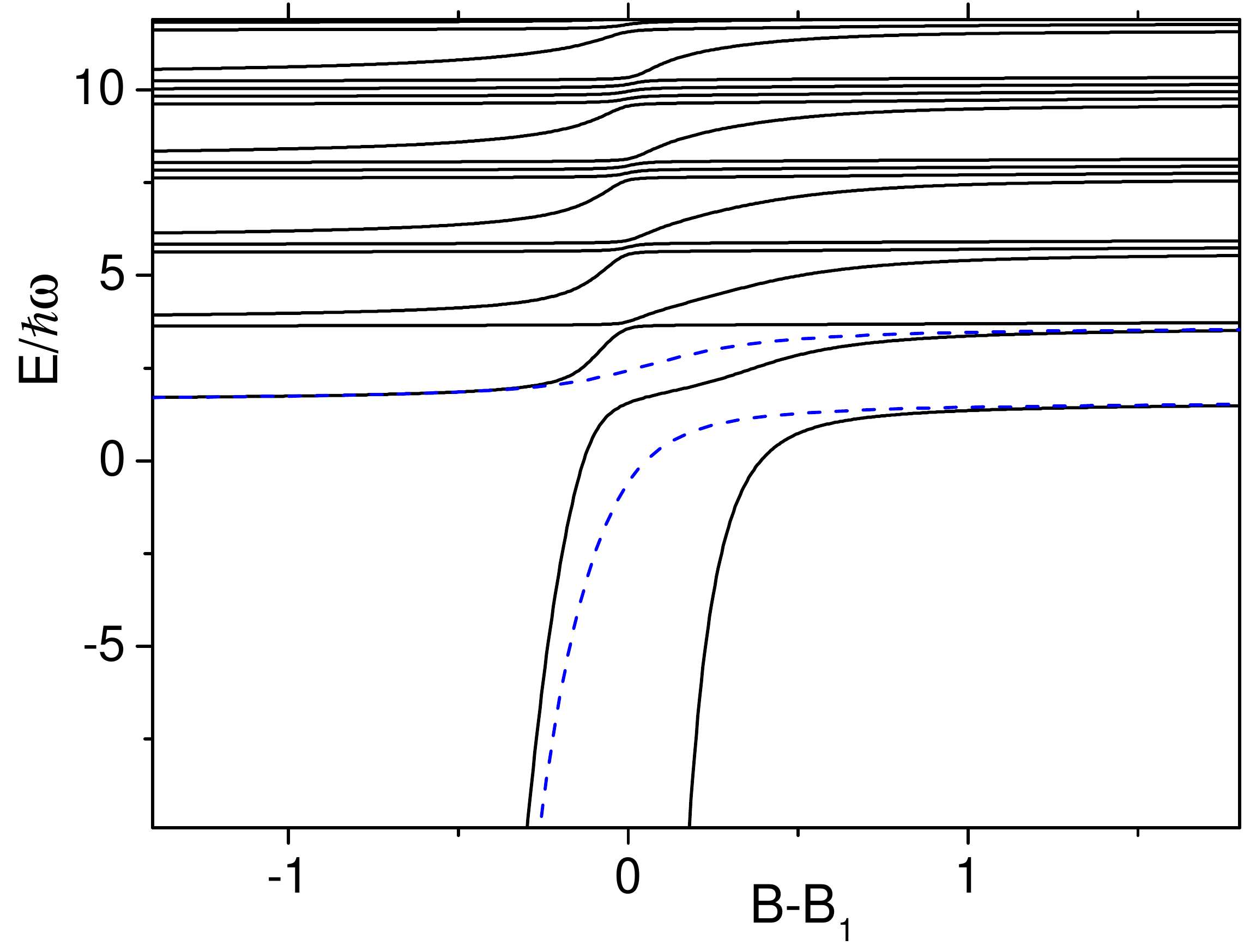}
\caption{\label{fig2}Role of the confinement strength. Two resonances in a slightly anisotropic trap ($\eta=1.1$) with widths $\Delta_1=2.5$G and $\Delta_2=0.5$G, separated from each other by $0.3$G. Left: $\omega=250$Hz. Right: $\omega=5$kHz. The blue dashed lines show the first two levels for the case in which there is only a single resonance present.}
\end{figure*}

Typically, the magnetoassociation process begins at magnetic fields higher than the resonance. The field is then slowly ramped across the resonance. The ground state of the system changes adiabatically from a pair of free atoms to a molecular bound state. The efficiency of the transfer can usually be well approximated by the Landau-Zener formula. For harmonic trap the probability of crossing the resonance diabatically and not producing a molecule reads~\cite{Julienne2004,Covey2015} $P=\mathrm{exp}\left(-\frac{4\sqrt{3}\omega a_{bg}\Delta}{a_{ho}dB/dt}\right)$. Analyzing this formula leads to an observation that in general higher trapping frequencies lead to larger probability for adiabatic transfer. This can be intuitively understood by imaging that very tight trap forces the wave functions of the atoms to overlap and thus increases the probability density of forming the molecule. Note also that using harmonic oscillator energy units in eq.~\eqref{eq_anf} naturally introduces $\sqrt{\omega}$ into the coupling strength $g$. However, the presence of overlapping resonances can change the situation. If the resonances are separated (by this we mean that there exists a magnetic field range between them where the eigenstates are free atomic pairs with no contribution from the molecular states), it may be possible to control the ramp speed to cross one of them diabatically, and the other adiabatically. However, for high trapping frequencies both resonances are broadened and it may no longer be possible to do it. Figure~\ref{fig2} shows such a case for an exemplary case of two dysprosium atoms with two resonances separated by 0.3 G. For a weak trapping potential of 250 Hz (left panel) the resonances are clearly separated and Landau-Zener theory should apply. As the confinement gets stronger (right panel), the states get mixed and the wave function contains contributions of both molecular states at once. One would then need to consider full multichannel dynamics to optimally control the magnetic field in order to produce only one kind of molecules.

\begin{figure*}
\includegraphics[width=0.4\textwidth]{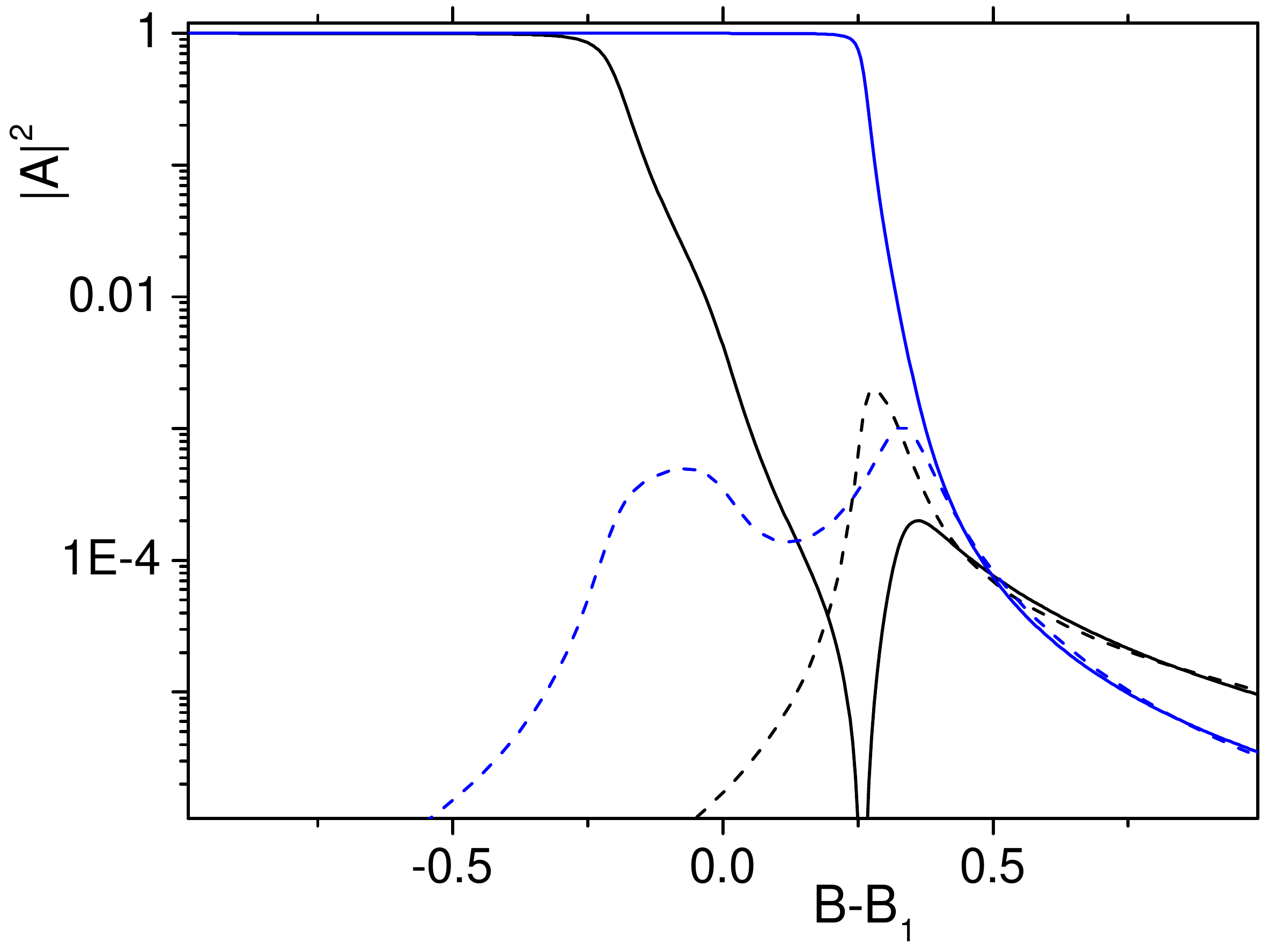}
\includegraphics[width=0.4\textwidth]{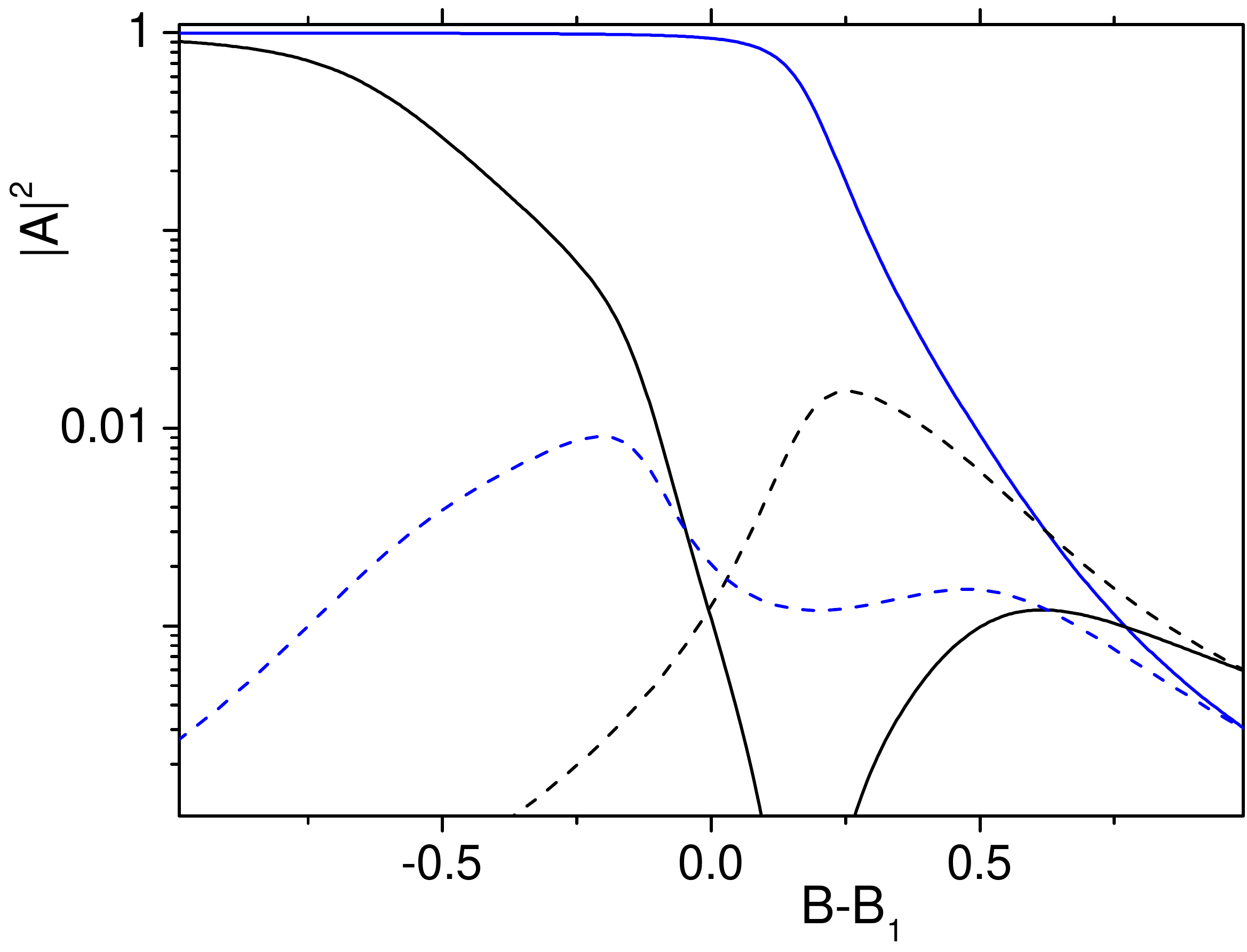}
\caption{\label{figc} Contributions of the two molecular bound states to the lowest levels for the same parameters as in Fig.~\ref{fig2}. Black lines show the state causing the wider resonance contributing to the lowest (dashed) and second lowest (straight) eigenstate. Blue lines describe the other molecular state contributing to the lowest (straight) and second lowest (dashed) eigenstate.}
\end{figure*}

For deeper understanding of this situation, Figure~\ref{figc} shows the contribution of the molecular bound states (measured as absolute square value of the respective coefficients in expansion~\eqref{eq_wavefharm}) to the lowest energy levels at different magnetic fields for weak ($\omega=1$kHz) and strong ($\omega=10$kHz) confinement for the same resonances as on Fig.~\ref{fig2}. For fields much below both resonances the lowest states are deeply bound and only contain pure molecular components. At the position of the first resonance the second lowest curve becomes mixed with other levels, while the lowest one is still deeply bound. Due to mixing of the two resonances, the straight black curve undergoes a minimum as the higher lying bound state starts to appear in the lowest state. For both resonances the contributions of different bound states become comparable for magnetic field slightly above the resonance positions, confirming the intuition provided by the energy level structure. Finally, at high fields both bound states retain a small contribution to the eigenstates which depends on the confinement strength. Figure~\ref{figc} also nicely demonstrates how tight confinement effectively makes the resonances wider.

\begin{figure*}
\includegraphics[width=0.4\textwidth]{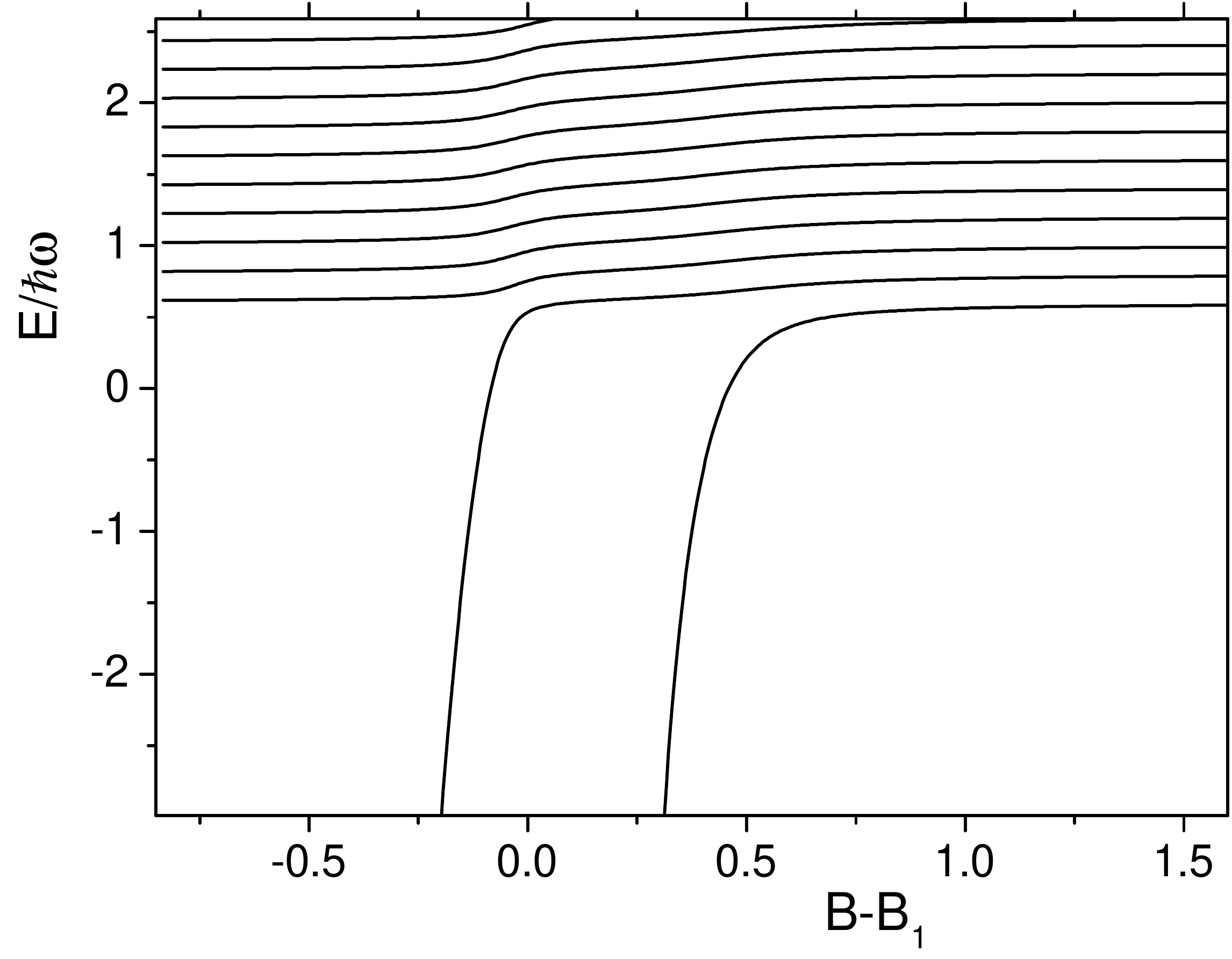}
\includegraphics[width=0.4\textwidth]{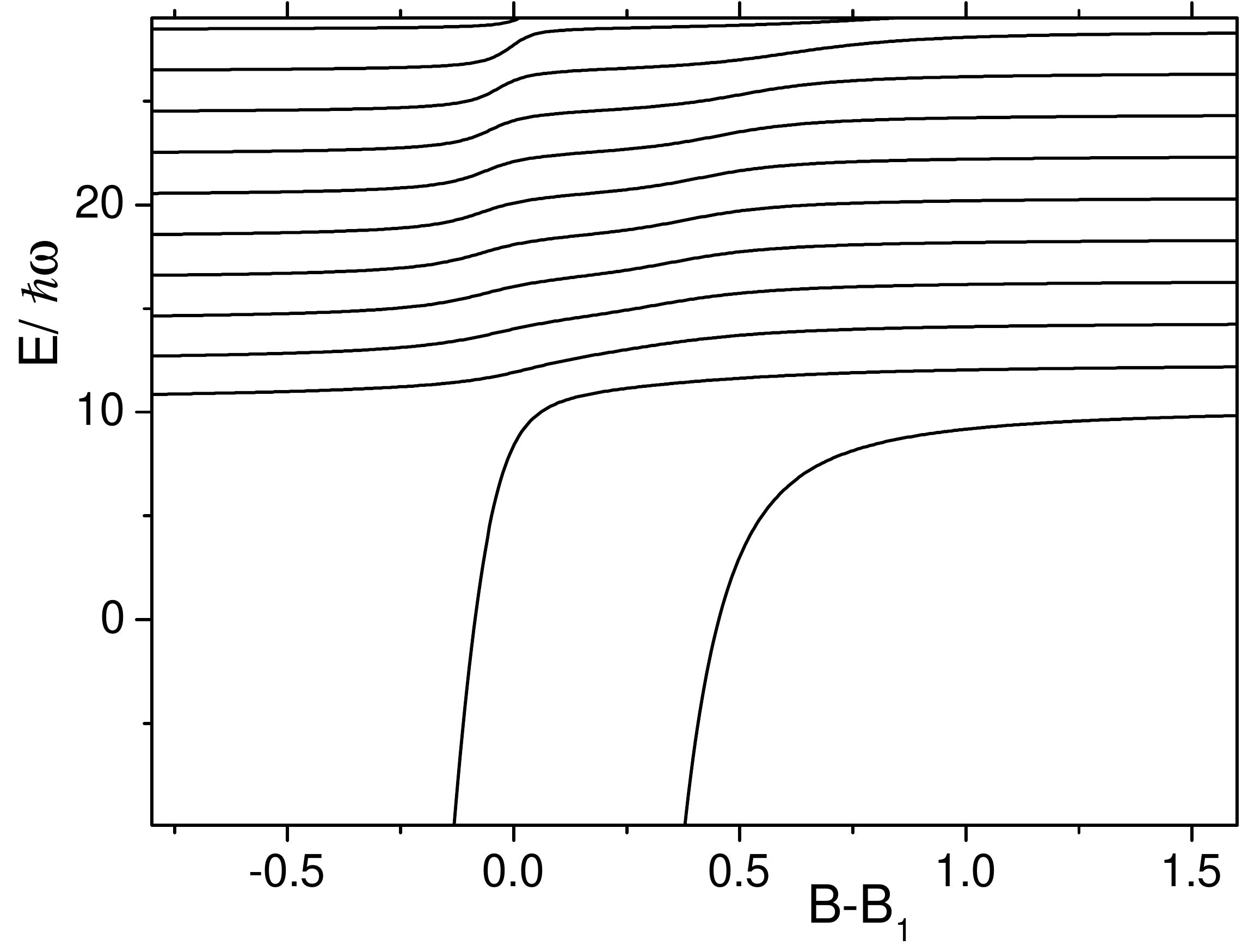}
\caption{\label{fig3}Role of the trap anisotropy. The same two resonances as in Figure~\ref{fig2}, but in a strongly anisotropic trap. Left: a quasi-2D (pancake) trap with $\omega=10$kHz $\eta=0.1$, right: quasi-1D (tube) trap with $\omega=1$kHz and $\eta=9.9$.}
\end{figure*}

It is also interesting to investigate the case of strongly anisotropic confinement, when the atoms are much more weakly trapped in one or two directions. It turns out that in this case the mean confinement is decisive. For the situation shown in Figure~\ref{fig3}, which depicts two cases of a pancake-shaped (left) quasi-2D trap and a tube-shaped (right) quasi-1D trap the mean confinement is strong and it does not seem realistic to separate the left resonance experimentally.

\begin{figure*}
\includegraphics[width=0.4\textwidth]{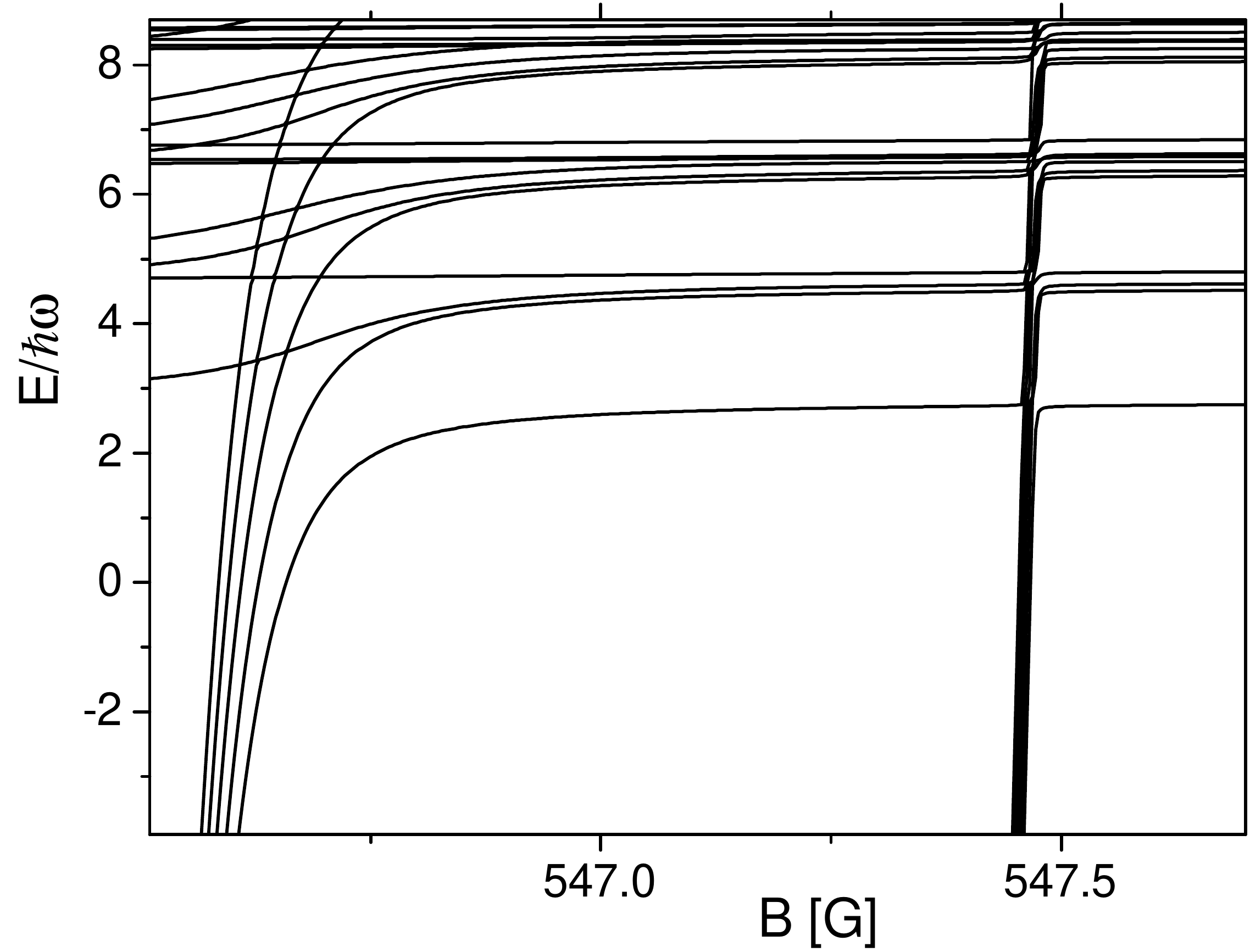}
\includegraphics[width=0.4\textwidth]{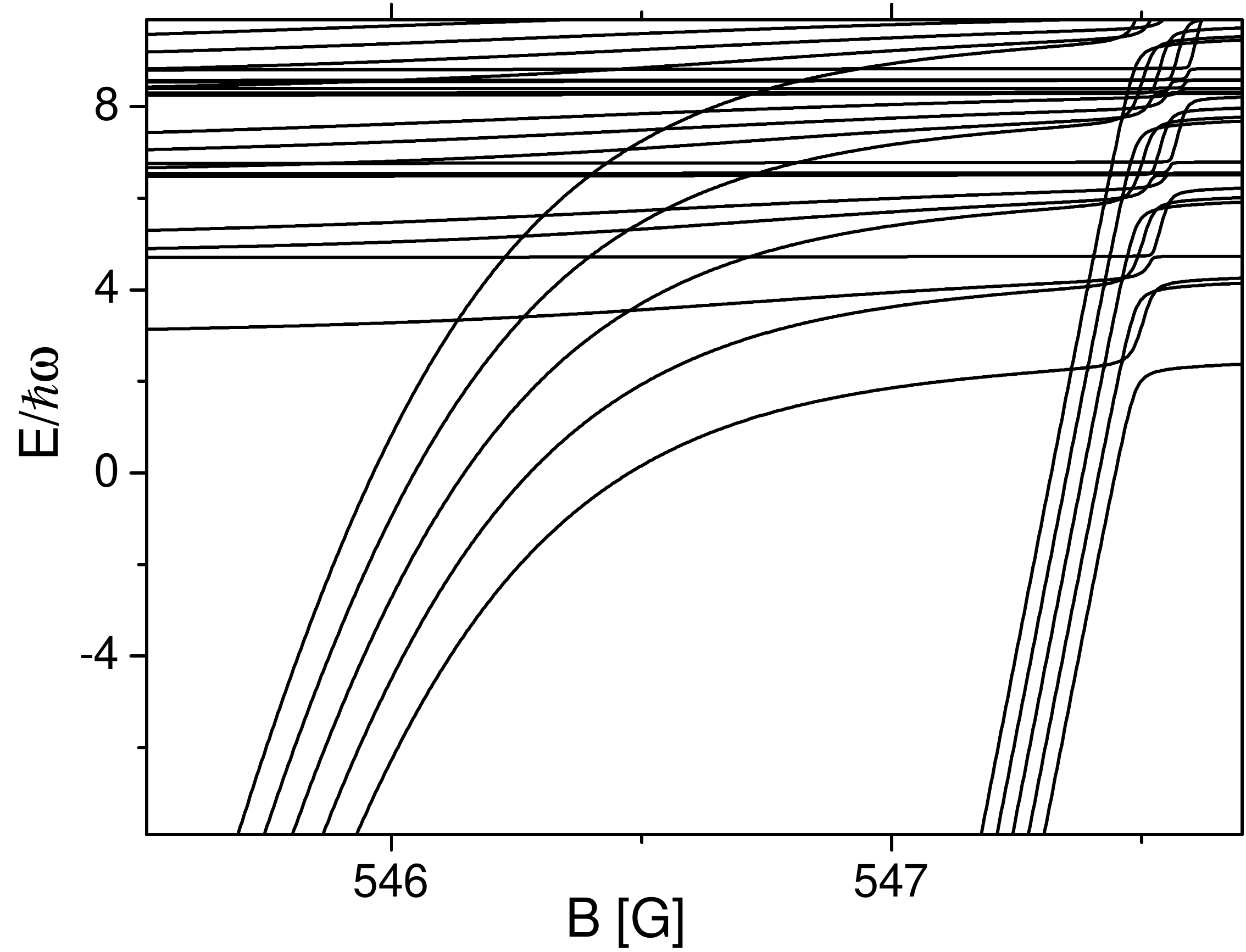}
\caption{\label{fig4}Energy levels for K-Rb resonance used in JILA experiment~\cite{Covey2015} for the trapping frequency $\omega=5$kHz (left) and 50kHz (right).}
\end{figure*}

So far, we have focused on the case of homonuclear molecules. For the heteronuclear case, such as KRb molecules studied in JILA~\cite{Ni2008,Moses2015,Covey2015}, one needs to take into account the coupling between the center of mass and relative motion $V_{rR}=C \mathbf{R}\cdot\mathbf{r}$, where $C=\mu(\omega_1^2-\omega_2^2)$~\cite{Sengstock}. This means that several molecular states in each closed channel have to be taken into account. We will focus on the spherically symmetric trap here, for which one can reduce the basis size by fixing the total angular momentum $J$~\cite{Bertelsen2007,Jachymski2013a}. As stated previously, the only needed extension is finding the open channel states. To this end, we set $J=0$ and use the following states as a basis
\beq
\Psi_{N\ell n}(\mathbf{R},\mathbf{r})=\sum_{m=-\ell}^{l}{\frac{(-1)^{\ell-m}}{\sqrt{2\ell+1}}\Phi_{N\ell m}(\mathbf{R})\phi_{n\ell(-m)}(\mathbf{r})},
\eeq
where $\Phi$ and $\phi$ are spherical harmonic oscillator states. The matrix element for the coupling term $V_{rR}$ can then be calculated analytically~\cite{Bertelsen2007}. The resulting open channel eigenstates contain components with different $\ell$, but only the $\ell=0$ part is coupled to the closed channels. They are also composed of multiple center of mass states, so each state can couple to multiple molecular states.

Figure~\ref{fig4} shows the calculated energy levels for the recently studied case of an $s$-wave resonance with $\Delta=3G$ in the KRb system overlapping with a narrow resonance coming from coupling to the $d$-wave bound state which is less than $0.01$G wide. The left panel shows moderate confinement strength, while the right one assumes very tight, but realistic trapping potential. For this case even at very strong confinement the resonances are quite well separated, which means that populating the $d$-wave molecular state can in principle be avoided.

\section{Discussion}
The model proposed in this paper bases on a number of simplifications. One of them is that the coupling between the channels can be described by Dirac delta. This requires that the characteristic interaction lengthscales are much smaller than the trap size. For van der Waals interactions the characteristic length $R_6=\left(2\mu C_6/\hbar^2\right)^{1/4}$ is usually of the order of a hundred Bohr radii. For very tight traps one may need to use more precise methods, such as multiscale quantum defect theory~\cite{Gao2007}.

\begin{figure}
\includegraphics[width=0.45\textwidth]{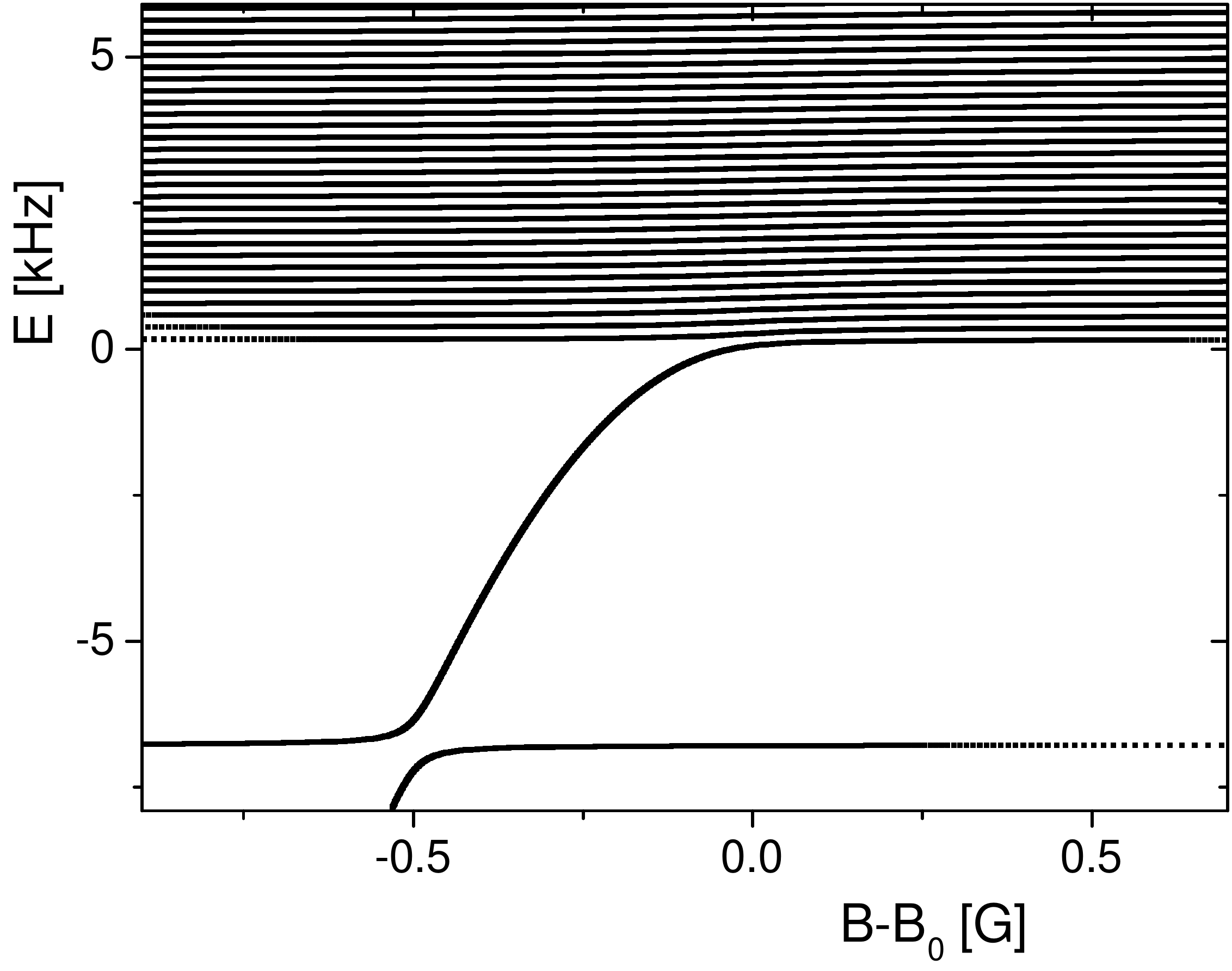}
\caption{\label{fig5}Energy levels for an exemplary Feshbach resonance with 0.5G width for cesium atoms placed in an isotropic harmonic trap with $\omega=100$Hz.}
\end{figure}

Another important approximation comes from neglecting the background interactions in the open channel. This allows us to use very simple channel wavefunctions, associated only with the trapping potential. The consequence is that the open channel does not by itself support any molecular bound states. As long as the first bound state associated with the background interaction is fairly deeply bound, we can safely neglect it in the calculation, as we are anyway only interested in the weakly bound states. However, for systems such as cesium, where the background scattering length is around 10$R_6$, the first bound state is located quite close to the threshold. For high trapping frequencies $E_b/\hbar\omega$ can become a small number and the energy levels will be strongly affected. This is illustrated in Figure~\ref{fig5} which shows the situation for cesium atoms placed in a trap with $\omega=100$Hz, computed using the analytical single-channel method~\cite{Idziaszek2006}. Already for this case the open channel bound state is non-negligible. In such a case one needs to explicitly add the interaction term to the open channel in eq.~\eqref{eq:ham} to reproduce the bound state. This makes the calculations more involved.

Recent experiment demonstrated association of dimers composed from erbium atoms~\cite{Frisch2015}. Lanthanide atoms are characterized by strong dipole-dipole interactions and very dense Feshbach spectra. It seems very challenging to construct a realistic description of such a system in a tight trap. Neglecting the dipolar part of the interaction and focusing only on resonances, as done in this work, seems to be a crude approximation. However, the main problem with this approach is the same as for cesium; namely, the lack of open channel bound states. Adding a properly designed pseudopotential to the open channel~\cite{Kanjilal2007} would again improve the applicability of the model.

The model presented here works well for describing multiple resonances, both wide and narrow, in tight nonseparable and anisotropic traps. If in addition strong background interactions are present or if the trap is very anharmonic, it might be a better idea to set up a full numerical multichannel model with realistic interactions and couplings instead of using the pseudopotential.

\section{Conclusions}
We proposed a method for computing the energy levels and eigenstates for two atoms in an external trap in the presence of multiple overlapping Feshbach resonances and applied it to the common experimental case of anisotropic harmonic confinement. The model is applicable to a number of systems which are currently being investigated in various laboratories. In particular, it allows for precise description of heteronuclear resonances. We analyzed the role of the trapping frequency in magnetoassociation of ultracold molecules and showed that although for a single resonance it is generally good to work at the strongest possible confinement, overlapping resonances may become impossible to separate as $\omega$ grows. It is possible to use our model to calculate the dynamics of atoms for an experimentally realistic case, which can be useful in further optimization of the molecule production process.

It would be interesting to analyze the impact of other types of anharmonicities on the energy level structure. In particular, shallow optical lattices~\cite{Buchler,Stecher2011,Sala2013} or optical tweezer potentials~\cite{Wall2015,Kaufman2015} can be readily studied with the method presented in this paper. One can expect that anharmonic terms can complicate the level structure for these cases. While such systems are definitely not optimal for the purpose of molecule production, they have huge potential for quantum simulation and computation, and Feshbach resonances will definitely be used there to control the interaction strength.

The author is grateful to Przemek Bienias for critical reading of the manuscript. This work was supported by the Foundation for Polish Science within the START program, Alexander von Humboldt Foundation and National Science Center project 2014/14/M/ST2/00015.

\bibliography{Allrefs}

\begin{thebibliography}{51}
\bibitem [{Gau()}]{GaussNote}%
  \BibitemOpen
  \href@noop {} {}\bibinfo {note} {Units of gauss rather than tesla, the
  accepted SI unit for the magnetic field, have been used in this paper to
  conform to the conventional usage in this field of physics.}\BibitemShut
  {Stop}%
\end{thebibliography}
\end{document}